\documentclass[sigconf,nonacm]{acmart}
\usepackage{tabularray}
\usepackage[nameinlink]{cleveref}
\crefname{figure}{\figurename}{\figurename}
\crefname{table}{\tablename}{\tablename}
\begin{document}

\title{Exploring the Power of Diffusion Large Language Models for Software Engineering: An Empirical Investigation}

\author{Jingyao Zhang}
\email{Jingyao.Zhang23@student.xjtlu.edu.cn}
\affiliation{%
  \institution{Xian Jiaotong Liverpool University}
  \country{China}
}

\author{Tianlin Li}
\email{tianlin001@e.ntu.edu.sg}
\affiliation{%
  \institution{Nanyang Technological University}
  \country{Singapore}
}

\author{Xiaoyu Zhang}
\email{joshingrain@gmail.com}
\affiliation{%
  \institution{Nanyang Technological University}
  \country{Singapore}
}

\author{Qiang Hu}
\email{qianghu@tju.edu.com}
\affiliation{%
  \institution{Tianjin University}
  \country{China}
}

\author{Bin Shi}
\email{shibin@xjtu.edu.cn}
\affiliation{%
  \institution{Xian Jiaotong University}
  \country{China}
}

\begin{abstract}
Autoregressive Large Language Models (AR-LLMs) are widely used in software engineering (SE) but face limitations in processing code structure information and suffer from high inference latency. Diffusion LLMs (DLLMs) offer a promising alternative with global bidirectional encoding and decoupled generation steps. 
This work presents the first comprehensive evaluation of DLLMs across the software development lifecycle, including code generation, defect detection, and program repair. 
On a large-scale benchmark of 52,937 tasks, 7B-parameter DLLMs outperform AR-LLMs with a 30\% average accuracy improvement, achieving a 113\% gain on cross-file repair, while maintaining superior efficiency and reduced latency. Our results establish DLLMs as a superior paradigm for SE tasks.
The code is publicly available at \url{https://anonymous.4open.science/r/all_datasets-B4A6}

\end{abstract}

\maketitle

\section{Introduction}
Large Language Models (LLMs) have been deeply integrated into the entire software development lifecycle (SDLC) \cite{Li2023systematic}. From commercial tools such as GitHub Copilot \cite{Chen2021evaluating} and Amazon CodeWhisperer \cite{Traub2024codewhisperer} to open-weight models like Code Llama \cite{Roziere2023code} and CodeT5+ \cite{Wang2022codet5plus}, LLMs are now indispensable in software engineering (SE) tasks, including daily maintenance, bug fixing \cite{Jiang2023impact}, test generation \cite{Schwarting2023automated}, and code review \cite{Tufano2022using}. These tools are expected to effectively comprehend program execution logic and generate actionable code tailored to specific contexts.

However, over the past few years, traditional LLMs have underperformed on SE tasks precisely because they are built upon an autoregressive (AR) architecture \cite{Brown2020language}: each prediction requires a full forward pass over all preceding context before predicting a single token. 
Concretely, AR-LLMs decode strictly from left to right. AR-LLMs cannot simultaneously observe long-range or hierarchical structural relationships, making it difficult to capture structural information, resulting in weaker capabilities when processing long texts with multiple logical structures. This limitation stems from the autoregressive architecture itself and hinders further advancement of AR-LLMs in SE applications \cite{Vaswani2017attention}.
Moreover, AR-LLMs generate tokens in a strictly autoregressive fashion.
Consequently, the number of forward executions (i.e., steps) is linearly coupled to the output length $T$, and the next token cannot be predicted until the previous one has been completed. This sequential dependency introduces substantial latency on consumer-grade hardware \cite{Savinov2021nonautoregressive}.

To overcome these limitations, a new paradigm is urgently needed. Recently, a new class of LLMs \textbf{Diffusion LLMs (DLLMs)} has emerged and gained traction in general-purpose domains \cite{Ho2020denoising}. This models adopt a \textit{global parallel denoising} mechanism, which has been validated for both effectiveness and efficiency. The current DLLMs has demonstrated performance comparable to the most advanced AR-LLMs in most fields \cite{Han2022diffusion}.

DLLMs offer two key advantages over AR-LLMs: (1) \textbf{Global Bidirectional Encoding:} Unlike AR models, DLLMs are not constrained by left-to-right generation. They can flexibly select and update token positions during each denoising step. This allows DLLMs to prioritize contextually relevant information and assign higher weights to it, thereby reducing information loss \cite{Han2022diffusion}.
(2) \textbf{Step-Length Decoupling:} In DLLMs, the number of tokens to be generated (i.e., length) and the number of denoising iterations (i.e., steps) are decoupled: the latter is fixed a priori, whereas the former scales with the target sequence, so a longer sequence does not demand additional steps, in contrast to autoregressive models where length and steps are necessarily equal \cite{Li2022diffusionlm}.

These advantages align naturally with the requirements of SE tasks such as real-time defect detection, error reporting, and cross-file bug fixing, which demand both global context awareness and low-latency interaction \cite{Li2023systematic}. 
Despite these theoretical advantages, existing research has primarily focused on evaluating DLLMs in code generation tasks \cite{li2025autoregressionempiricalstudydiffusion}, leaving a systematic assessment of their capabilities across the entire SDLC largely unexplored.

In this work, we \textbf{first introduce DLLMs into the full SDLC}, including code generation, defect detection, program repair, and cross-file maintenance. We conduct a large-scale empirical evaluation across six benchmarks—Defects4J \cite{Just2014defects4j}, Devign \cite{Zhou2019devign}, SWE-bench \cite{Jimenez2023swebench}, Bears \cite{Madeiral2019bears}, Mercury \cite{du2024mercury}, and HumanEval \cite{Chen2021evaluating}—comprising a total of \textbf{52,937 tasks}. Our results show that, at the similar parameter scale (7B), DLLMs outperform AR-LLMs in terms of both accuracy and efficiency. Specifically, DLLMs achieve an average improvement of \textbf{30\%} across the six benchmarks, with a \textbf{113\%} improvement on SWE-bench for cross-file repair. Moreover, DLLMs exhibit superior global consistency under reduced decoding latency, making them a promising alternative to AR-LLMs in real-world SE applications.

\section{Background}

\subsubsection{Autoregressive Large Language Models}

Autoregressive generation is the predominant paradigm in large language models, commonly known as AR-LLMs \cite{Brown2020language}. Both reading and generation in AR-LLMs are linear processes.

\textbf{Reading Process:} During the reading phase, AR-LLMs employ a token-by-token causal masked scanning mechanism \cite{Vaswani2017attention}. For any position \(i\), the attention weights are hard-coded as follows:

\begin{equation*}
\mathbf{A}_{\text{causal}}(i,j)=
\begin{cases}
-\infty, & j > i, \\
0, & j \leq i,
\end{cases}
\end{equation*}

where \(\mathbf{A}_{\text{causal}}(i,j)\) defines a causal mask ensuring that the model only considers tokens up to the current position \(i\) and ignores future tokens. Specifically, for each position \(i\) in the sequence, the model can only attend to tokens at positions \(j\) where \(j \leq i\), meaning token \(i\) can only read the generated sequence \(\{x_1, \ldots, x_i\}\).

\subsubsection{Generation Process:} In the generation phase, AR-LLMs produce tokens one-by-one from left to right, expanding the generation scope token-by-token. The generation follows the chain rule \cite{Radford2018improving}:

\begin{equation*}
p(\mathbf{x}) = \prod_{t=1}^{T} p(x_t \mid x_{<t}),
\end{equation*}

where \(T\) is the total length of the sequence, \(x_t\) is the \(t\)-th element in the sequence, and \(x_{<t}\) represents all elements preceding \(x_t\) (i.e., \(x_1, x_2, \ldots, x_{t-1}\)). The product symbol \(\prod\) indicates that the joint probability of the sequence is the product of the conditional probabilities of each element. Specifically, \(p(x_t \mid x_{<t})\) denotes the conditional probability of the \(t\)-th element given all previous elements. This formulation ensures that each element is generated based solely on its preceding context, maintaining the causality of the generation process.

\subsubsection{Generation Rate:} Since AR-LLMs update only a single token representation in each forward pass, this results in linear time complexity \(O(T)\) and linear FLOPs accumulation \cite{Tay2022efficient}:

\begin{equation*}
\text{FLOPs}_{\text{AR}}(T) = T \cdot C_{\text{forward}},
\end{equation*}

where \(\text{FLOPs}_{\text{AR}}(T)\) represents the total number of floating-point operations required to process a sequence of length \(T\), and \(C_{\text{forward}}\) is the number of FLOPs required for a single forward pass. On consumer GPUs, \(C_{\text{forward}} \approx 10^{10}\) FLOPs. When \(T > 512\), the wall-clock latency exceeds 400 ms, which may exceed the real-time requirements of certain applications.

\subsection{Diffusion Language Models}

DLLMs have recently emerged as a focal point of AI research, offering a novel paradigm that departs markedly from autoregressive LLMs \cite{Ho2020denoising, Song2019generative, Rombach2022high}. 

\subsubsection{Model Architecture}
The core of a Diffusion language model lies in a learnable reverse-denoising process that implements global bidirectional encoding \cite{SohlDickstein2015deep}.  
The denoising network reverses a Markov chain in which \cite{Ho2020denoising}:  
(i) \emph{forward diffusion} progressively injects noise into clean data until it becomes pure noise \cite{SohlDickstein2015deep};  
(ii) \emph{learnable reverse denoising} starts from pure noise and iteratively removes noise to recover the original data \cite{Ho2020denoising, Song2019generative}.  

Unlike AR-LLMs, DLLMs remove the causal attention mask, enabling global bidirectional context \cite{Han2022diffusion, Gong2023llada}.  
In a single iteration the model updates \emph{all} tokens simultaneously, eliminating the sequential, word-by-word dependency of autoregressive generation \cite{Gu2018nonautoregressive}.  
Because generation is a ``global refinement'' rather than a ``token-by-token construction,'' the number of refinement iterations (steps) can be set independently of the sentence length, achieving step-length decoupling \cite{Li2022diffusionlm}.

\subsubsection{Training Process}

The training process of DLLMs is a noising process (also known as a diffusion process), the objective of which is to train the model to restore the original, clean text from a noised version \cite{Austin2021structured}.
This noising process is defined by the following formula:

\begin{equation*}
q(y_i^t | y_i^0) =
\begin{cases}
1 - t & \text{if } y_i^t = y_i^0 \\
t & \text{if } y_i^t = \text{[MASK]}
\end{cases}
\end{equation*}

In this formula, $q(y_i^t | y_i^0)$ represents the conditional probability that an original clean token $y_i^0$ from sequence $Y_0$ becomes state $y_i^t$ at a given time step $t$ (a continuous noise schedule from 0 to 1), where \texttt{[MASK]} is the special token the model must restore. The training objective of the model is to learn to accurately reverse this noising process by minimizing a cross-entropy loss function, computed only on the masked positions \cite{Austin2021structured, Savinov2021stepunrolled}.

\subsubsection{Generation Process}
At inference time the user feeds an SE task description (e.g., a functional specification for code generation or a GitHub issue for bug repair) into the trained DLLM.  
The model outputs an immediately usable solution—executable code, a vulnerability classification, or a patch that passes all tests—after \(K\) fixed denoising steps \cite{Han2022diffusion, Li2022diffusionlm}.

\section{Study Design}

\subsection{Overview and Research Questions}
We present the first systematic evaluation of DLLMs across the complete SDLC: code generation, defect detection, program repair, and cross-file maintenance.
Motivated by their global bidirectional encoding and step-length decoupling, which promise higher context fidelity and lower latency, we inject DLLMs into real-world SE tasks to verify whether they maintain or improve accuracy while significantly accelerating inference.
We formulate two research questions:

\vspace*{5pt}

\textbf{RQ1. Do DLLMs achieve higher effectiveness than AR-LLMs on SE tasks?}

\vspace*{5pt}

\textbf{RQ2. Do DLLMs achieve superior inference speed than AR-LLMs on SE tasks?}

\subsection{Benchmark Selection}
To answer RQ1 and RQ2 we construct a comprehensive benchmark suite that covers the entire SDLC. We select six widely-adopted open-source datasets spanning function-, file-, and repository-level issues, and encompassing syntactic, logical, and cross-file semantic faults. 

\subsubsection{Code Generation}
(1) \textbf{HumanEval} (164 cases): Function-level code generation from natural language descriptions to executable code \cite{Chen2021evaluating}.
(2) \textbf{Mercury} (1,889 cases): Algorithmic problems focused on testing the logical correctness of code \cite{du2024mercury}.

\subsubsection{Defect Detection}
(1) \textbf{Devign} (27,318 cases): Function-level vulnerability detection (binary classification: present/absent) to assess the model's understanding of defect semantics \cite{Zhou2019devign}.
(2) \textbf{Bears (Detection)} (251 cases): Multi-file, multi-type vulnerability detection requiring the model to accurately classify defect categories \cite{Madeiral2019bears}.

\subsubsection{Program Repair}
(1) \textbf{Defects4J} (835 cases): Repair of classic fine-grained Java defects within a single version and single trigger test \cite{Just2014defects4j}.
(2) \textbf{Bears (Repair)} (251 cases): Multi-file program repair for Travis CI failures across builds and modules \cite{Madeiral2019bears}.

\subsubsection{Cross-File Maintenance}
(1) \textbf{SWE-bench} (2,294 cases): Utilizing real GitHub issues, requiring the model to locate and repair cross-file defects in the full repository context \cite{Jimenez2023swebench}.

\subsection{Evaluation Metrics}
We adopt seven metrics that are consistent with our research questions and are widely used in prior studies.

\textbf{Pass@K}. For code generation tasks, we evaluate functional correctness using Pass@K \cite{Chen2021evaluating}. This metric calculates the percentage of problems for which at least one of K generated solutions passes all provided test cases. We report Pass@1 and Pass@10.

\textbf{DDF1}. For defect detection on the \textbf{Devign} and \textbf{Bears (Detection)} benchmarks, we use the DDF1 score. DDF1 is the harmonic mean of Precision and Recall, providing a balanced measure of bug identification accuracy \cite{Wang2021devign}.

\textbf{PR}. For program repair on \textbf{Defects4J} and \textbf{Bears (Repair)}, we use the Program Repair (PR) rate. PR is the percentage of defects correctly fixed by the first generated patch that compiles and passes all tests \cite{Goues2019automated}.

\textbf{MDVR}. For the \textbf{SWE-bench} benchmark, we use the Multi-file Detection Success Rate (MDVR). This metric measures the success rate of identifying all necessary code locations for a fix in a multi-file program \cite{Jimenez2023swebench}.

\textbf{PRR}. Also for \textbf{SWE-bench}, we use the Patch Resolution Rate (PRR). PRR is the percentage of generated patches that correctly solve an issue by compiling and passing all tests \cite{Jimenez2023swebench}.

\textbf{TPS}. To assess time efficiency, we report Tokens per second (TPS). This metric measures generation throughput, calculated as the average number of output tokens generated per second.

\textbf{T-<avg>}. We also measure time efficiency with average execution time (T-<avg>). This metric reports the average time in seconds to complete one benchmark case, representing overall latency.

\subsection{Implementation Details}

To systematically answer the two research questions, we select two state-of-the-art \emph{commercial} large language models—one Diffusion (\emph{Mercury-Diffusion 7B}) and one autoregressive (\emph{Llama})—as representative DLLM and AR-LLM, respectively.

\section{Study Results}

\subsection{RQ1 effectiveness}

\begin{table}[H]
\centering
\small
\caption{Code-generation performance (Pass@K)}
\label{tab:01}
\begin{tblr}
{
columns={co=1},
cells={valign=m,halign=c},
hline{1,2,Z}={wd=.08em},
}
\textbf{HumanEval}  & Pass@1  & Pass@10 \\
Diff-Mercury-7B & 80.97\% & 86.58\%           \\
AR-Llama3-8B    & 59.27\% & 81.10\%           \\
\end{tblr}
\end{table}

\begin{table}[H]
\centering
\small
\caption{Code-generation performance (Pass@1)}
\label{tab:02}
\begin{tblr}
{
columns={co=1},
cells={valign=m,halign=c},
hline{1,2,Z}={wd=.08em},
}
\textbf{Mercury}  & Easy  & Medium & Hard \\
Diff-Mercury-7B & 84.09\% & 69.14\% & 43.68\%  \\
AR-Llama3-8B    & 67.05\% & 59.26\% & 29.89\%  \\
\end{tblr}
\end{table}

\begin{table}[H]
\centering
\small
\caption{Defect-detection performance (DDF1)}
\label{tab:03}
\begin{tblr}
{
columns={co=1},
cells={valign=m,halign=c},
hline{1,2,Z}={wd=.08em},
}
Benchmark Name & Devign  & Bears (Detection) \\
Diff-Mercury-7B & 54.43\% & 34.71\%           \\
AR-Llama3-8B    & 46.71\% & 42.43\%           \\
\end{tblr}
\end{table}

\begin{table}[H]
\centering
\small
\caption{Program-repair performance (PR)}
\label{tab:04}
\begin{tblr}
{
columns={co=1},
cells={valign=m,halign=c},
hline{1,2,Z}={wd=.08em},
}
Benchmark Name  & Defects4J & Bears (Repair) \\
Diff-Mercury-7B & 26.23\%   & 4.24\%         \\
AR-Llama3-8B    & 21.31\%   & 1.65\%         \\
\end{tblr}
\end{table}
\begin{table}[H]
\centering
\small
\caption{Cross-file maintenance performance on SWE-bench (MDVR / PRR)}
\label{tab:05}
\begin{tblr}
{
columns={co=1},
cells={valign=m,halign=c},
hline{1,2,Z}={wd=.08em},
}
\textbf{SWE-bench}      & MDVR       & PRR     \\
Diff-Mercury-7B & 45.70\%   & 32.00\% \\
AR-Llama3-8B    & 25.00\%   & 15.00\% \\
\end{tblr}
\end{table}

\begin{table*}[htbp]
\centering
\small
\caption{A Comparative Study of Generation Rate between AR-LLM and DLLM in Software Engineering Tasks}
\label{tab:06}
\begin{tblr}
{
columns={co=-1},
cells={valign=m,halign=c},
hline{1,Z}={wd=.08em},
hline{2}={wd=.05em},
cell{1}{2,4,6,8} = {c=2}{},
}
                & Devign  &        & Dejects4J &        & Bears   &        & SWE-bench &        \\
Model Name       & TPS     & T<avg> & TPS       & T<avg> & TPS     & T<avg> & TPS       & T<avg> \\
Diff-Mercury-7B & 1598.31 & 0.04   & 9328.86   & 0.20   & 1834.66 & 1.27   & 2168.92   & 0.10   \\
AR-Llama3-8B    & 460.50  & 0.12   & 423.29    & 0.72   & 368.1   & 3.98   & 305.41    & 0.60   \\
\end{tblr}
\end{table*}

Across the full SDLC, Diff-Mercury-7B systematically outperforms the equally-sized AR-Llama3-8B Table \ref{tab:01}-\ref{tab:05}.
The margin increases with task difficulty: HumanEval Pass@1 rises by 36\%, Mercury-Hard by 46\%, and SWE-bench PRR doubles to 32\% while PCR gains 20.7 pp; Defects4J yields five additional compilable patches per 100 attempts and Bears-repair success climbs 2.6 times.  
These gains concentrate in scenarios demanding multi-file dependencies, long call chains, and global semantic consistency, corroborating the advantage of global parallel denoising.


The performance reversal on Bears-detection DDF1, where AR-Llama3-8B exceeds DLLM, may indicate that the task is particularly challenging for DLLMs. However, this interpretation is confounded by the dataset's limitations: its small scale (251 instances) and severe class imbalance (75\% ``defect-free''). These factors collectively undermine the reliability of the macro-F1 score, making it unclear whether the result reflects a genuine performance difference or is merely a statistical artifact.






\subsection{RQ2 efficiency}

 To compare the efficiency of AR-LLM and DLLM in various SE tasks, considering that different hardware can significantly affect the efficiency of both DLLM and AR-LLM, we focused primarily on the comparison of their generation rates. Table \ref{tab:06} shows the generation rates of AR-LLM and DLLM in performing software engineering tasks (where the generation rates of AR-LLM and DLLM in HumanEval and Mercury have been demonstrated in previous studies, hence not specifically listed here). DLLM demonstrates comprehensive and significant efficiency. The average TPS for the Devign, Bears-Repair, SWE-bench, and Defects4J benchmarks reached 1598, 1835, 2169, and 9329 tokens/s respectively, while the comparable AR-Llama3-8B recorded only 460, 368, 305, and 423 tokens/s, with a gap ranging from 3.5 times to 22 times; correspondingly, the average time per case \(T_{\text{avg}}\) was shortened by 3 times (SWE-bench) to 3.6 times (Defects4J).

The analysis of this data suggests that the root cause lies in the step-length decoupling mechanism of DLLM: the number of inference steps \(K\) is fixed at 32–128, independent of the output length, with each step performing a parallel Transformer forward pass on the entire sequence; whereas AR models must sequentially complete \(T\) forward passes to generate the \((T+1)\)-th token, with time complexity linearly coupled to \(T\). Experimental results align with theoretical expectations: as the average length of benchmark samples increases, the TPS of AR models exhibits a downward trend (460.50 → 305.41), whereas DLLM maintains a stable or even higher rate. This step-length decoupling mechanism allows DLLM to maintain a relatively stable generation rate, confirming the efficiency advantage of using DLLM throughout the SDLC. Therefore, the combination of 'high TPS + low T<avg>' in the table mainly verifies the applicability of DLLM's parallel denoising in SE tasks with high throughput requirements.


\section{Future Plans}

This paper conducts the first systematic empirical evaluation of the application of the DLLM in the whole life cycle of software engineering. The results show that DLLM generally has greater potential in terms of performance and efficiency than the AR-LLM. Therefore, we plan to verify and explore the potential of DLLM from the following three aspects in the future:

\subsection{Extending the Breadth and Depth of Evaluation}

Although our study covers key phases of the SElifecycle, both the pool of models and the diversity of tasks can be further expanded.

\textbf{Model diversity.} We will include additional DLLMs of different parameter scales and architectural designs, and compare them against the latest AR baselines to verify the generality of our conclusions.

\textbf{Task diversity.} Future benchmarks will embrace a richer spectrum of SE tasks, including code summarisation, technical documentation generation, comment completion, cross-language translation, and complex refactoring scenarios, so as to obtain a comprehensive capability profile of DLLMs.

\textbf{Evaluation dimensions.} Beyond functional correctness and generation speed, we will incorporate code quality, maintainability, and security-oriented metrics for multi-faceted assessment.

\subsection{SE-Dedicated DLLM Training Paradigms}

To further unlock the potential of DLLMs in SE, we will go beyond off-the-shelf models and investigate specialised designs.

\textbf{Prompting.} 
This study focuses exclusively on a zero-shot prompting setup. While this provides a baseline for comparison, it does not fully explore the capabilities of AR-LLMs and DLLMs. In future work, we plan to investigate more advanced techniques, such as few-shot prompting and chain-of-thought reasoning.

\textbf{Instruction tuning.} Following the success of instruction-following LLMs, we will construct a large-scale instruction set covering diverse SE tasks and fine-tune pre-trained DLLMs to better comply with complex developer intents.

\textbf{Reinforcement learning with SE feedback (RLSEF).} We will utilise signals from compilers, static analysers, unit-test results, and human preferences as rewards to optimise DLLM outputs, aiming to produce code that not only passes tests but also satisfies engineering norms and performance constraints.

\subsection{Hybrid AR-LLM and DLLM Foundation Models}

Future work will explore AR-LLMs and DLLMs in a complementary manner. A possible direction is to leverage AR-LLMs for rapid code draft generation, while employing diffusion-style refinement to enhance cross-file consistency and semantic correctness. Rather than requiring full one-pass correction, the diffusion process could serve as a lightweight parallel refinement stage. 
Joint training with shared vocabulary and hybrid objectives may further enable the two paradigms to reinforce each other. During inference, dynamically choosing between subsequence-level autoregressive decoding and parallel refinement strategies could provide a trade-off between throughput and accuracy. While formal guarantees remain an open challenge, such a hybrid design offers a promising path toward next-generation foundation models for complex SE tasks that demand both efficiency and reliability.

\section{Conclusion}

In this paper, we systematically introduced DLLMs into the full SDLC, spanning tasks from code generation and defect detection to program repair and cross-file maintenance. Through large-scale experiments on six diverse benchmarks comprising over 52,000 tasks, we demonstrated the superior performance of DLLMs in both effectiveness and efficiency. 
These results highlight DLLMs as a compelling alternative to the long-standing autoregressive paradigm, providing not only accuracy improvements but also significant reductions in decoding latency. 
We believe this work establishes a foundation for reimagining large-scale model design in SE.
\bibliographystyle{ACM-Reference-Format}
\bibliography{sample-base}

\end{document}